\documentstyle[12pt,fleqn]{article}
\def\limh{\lim_{H\rightarrow 0}}
\def\limt{\lim_{T\rightarrow 0}}
\def\gamh{\frac{N(N-1)}{12J}\left[1+\frac{\sum_{j=1}^{N-1}
          G_{+}^{(N,m,j)}(i\frac{2\pi}{N})}{N-1}\right]}
\def\gamt{\frac{N(N^2-1)m}{6J(m+N)}}
\def\gam{\gamma^{[N,m]}}
\def\sus{\frac{\chi(H)}{\chi(0)}}

\def\sums{\sum_{s=1}^{N-1}}
\def\prodl{\prod_{{l=0,l\not=j}}^{m}}
\def\epj{\varepsilon_{j,m}(\lambda)}
\def\ep{\varepsilon_{r,m}(\lambda)}

\def\intinf{\int_{-\infty}^{\infty}}
\def\epsnt{\varepsilon_{j,m}^{(0)}}
\def\epst{\varepsilon_{j,m}^{(2)}}
\def\deps{\left| \frac{d\varepsilon_{j,m}^{(0)}}{d\lambda}\right|}

\begin{document}
\baselineskip=24pt

% title and abstract

\begin{titlepage}
\begin{center}{\LARGE Anomalous low-temperature specific heat \\
of the antiferromagnetic SU(N) Heisenberg model in a field}
\end{center}
\vskip 24pt
\begin{center}{K. Lee\\
Department of Physics,\\ Ewha Woman's University,
Seoul 120-750, Korea}

\end{center}
\vskip 48pt
\begin{abstract}

We discuss the low-temperature specific heat of the integrable SU(N)-
invariant Heisenberg model in one dimension with degrees of freedom in
the symmetric rank-$m$ tensor representation, especially for the
antiferromagnetic coupling. It is known that the linear specific heat
coefficient $\gam$ is a function of a field which breaks the SU(N)
invariance of the internal degrees of freedom.
We calculate the $\gam$ in a zero field and in a small field.  The in-field
$\gam$ is less than the zero-field $\gam$ as expected since the entropy is
reduced in the ordered system.  The zero-field $\gam$ is the same as the
one obtained by the prediction of the critical behavior of 1D quantum spin
systems via conformal field theory. This extends the previous results for
N=2 to an arbitrary N.

\end{abstract}
\end{titlepage}

% main text file
\setcounter{page}{2}

The Bethe-ansatz method yields singular behavior in numerous other models
with SU(N) symmetry, for instance, the N-component 1D fermion gas
interacting via a $\delta$-functional repulsion \cite{sut1}-\cite{sch1},
the Gross-Neveu model \cite{al}, the N-component supersymmetric 1D t-j
model \cite{ls1}, and the N-degenerate 1D Hubbard chain \cite{ls2, sch2}.
In particular, the logarithmic singularities in zero-temperature
susceptibility to a field are common to these models. For a small Zeeman
splitting $H$ and for arbitrary $N$ and $m$ \cite{sch3},
\begin{equation}
\sus=\left( 1+\frac{m}{N|ln H|}-\frac{m^2 ln|NlnH|}
{N^2 ln^2H}+\cdots \, \right)\, .  \label{sus}
\end{equation}
The nonanalytic field dependence of the susceptibility suggests a singular
behavior in the low-temperature specific heat, $C=\gamma T$, as $H$ and
$T$ tend to zero \cite{sch3}-\cite{lee}, i.e.,
\begin{equation}
	\limh\limt\gam\neq\limt\limh\gam .        \label{sh}
\end{equation}

In this letter we calculate the linear specific heat coefficient $\gam$
in the limit of $T\rightarrow 0$ with and without the SU(N)-invariance
breaking field. Our main result is that for a small magnetic field $H$,
\begin{eqnarray}
	\limh\limt \gam & = & \gamh \nonumber \\
			 &\not=&\limt\limh \gam =\gamt,  \label{gam}
\end{eqnarray}
where
\begin{eqnarray}
G_{+}^{(N,m,j)}(\omega)&=&\left(\frac{-i\omega+0}{a_{N,m,j}}\right)
^{i\frac{m\omega}{2\pi}}\sqrt{\frac{j(N-j)}{mN}}\frac{\Gamma
(1-i\frac{N\omega}{2\pi})\Gamma(1-i\frac{m\omega}{2\pi})}
{\Gamma(1-i\frac{j\omega}{2\pi})\Gamma(1-i\frac{(N-j)\omega}{2\pi})},
\label{g+}
\\ [12pt]
a_{N,m,j}&=&\frac{2\pi e j^{\frac{j}{m}}(N-j)^{\frac{N-j}{m}}}
{mN^{\frac{N}{m}}} , \nonumber
\end{eqnarray}
\vspace{12pt}
with $\Gamma$ being the gamma function.

The generalized SU(N) Heisenberg model in one dimension with degrees of
freedom in the symmetric rank-$m$ tensor representation is given by
\cite{aj, joh}
\begin{equation}
H^{[N,m]}=J\sum_{k=1}^{N_a}I_{k,k+1}^{[N,m]},  \label{ham}
\end{equation}
with
\begin{eqnarray}
I_{k,k+1}^{[N,m]} & = & \sum_{j=1}^{m}\sum_{s=0}^{j-1}(m-s)^{-1}
		 \prodl\alpha_{jl}^{m}({\bf A}_{k}\otimes{\bf A}_{k+1}
+\beta_{l}^{[N,m]}), \nonumber  \\
\alpha_{jl}^{m} & \equiv & [\frac{1}{2}j(j-1)-\frac{1}{2}l(l-1)-m(j-l)]
^{-1}, \nonumber \\
\beta_{l}^{[N,m]} & \equiv & l(2m+1-l)-m^{2}(N-1)/N,  \nonumber
\end{eqnarray}
where $J>0$ is the antiferromagnetic coupling constant, $k$ labels the
$N_{a}$ lattice sites in the chain and ${{\bf A}_{k}}$ denotes the
generators of the symmetric rank-$m$ tensor representation, playing the
role of spin operators.

The Bethe's ansatz method applied to this model yields $N-1$ sets of
rapidities and all rapidities within a given set are different. Hence
rapidities obey Fermi statistics. In general, these rapidities are
complex and form so called, strings for a very large system such that
the center of mass of each string becomes real. At finite $T$ the length
$n$ of strings could be infinitely long. In the thermodynamic limit we
define the densities of states for occupied rapidities as $\sigma_{r,n}
(\lambda)$ and for unoccupied ones as $\sigma_{r,n}^h (\lambda)$.
These densities then satisfy the following integral equations with
$n=1,\cdots,\infty$,

\begin{equation}
\sigma_{r,n}(\lambda)+\sigma_{r,n}^h(\lambda)-\sums P_{r,s}^N*
(\sigma_{s,n+1}^h+\sigma_{s,n-1}^h
			-\sigma_{s-1,n}^h-\sigma_{s+1,n}^h)
		     =F_r^N(\lambda)\delta_{nm},
\label{den}
\end{equation}
where $\sigma_{0,n}\equiv\sigma_{N,n}\equiv 0$ for all $n$, the
asterisk denotes a convolution, $P_{r,s}^N$ is the Fourier transform of
\begin{equation}
\tilde{P}_{r,s}^N(\omega)=\frac{\sinh(min(r,s)\frac{\omega}{2})
\sinh([N-max(r,s)]\frac{\omega}{2})}
		       {\sinh(\frac{\omega}{2})\sinh(N\frac{\omega}{2})},
\label{prs}
\end{equation}
and
\begin{equation}
F_r^N(\lambda)=\frac{1}{N}\frac{\sin(\pi\frac{N-r}{N})}{\cosh (
\frac{2\pi}{N}\lambda)+\cos(\pi\frac{N-r}{N})} \, . \label{frn}
\end{equation}

It is usual to introduce the thermodynamic energy potentials
$\varepsilon_{r,n}(\lambda)$ such that $\sigma_{r,n}^h(\lambda)/
\sigma_{r,n}(\lambda)=exp(\varepsilon_{r,n}(\lambda)/T)$. Hence the
strings with the same lengths in each family  form energy
bands denoted by $\varepsilon_{r,n}(\lambda)$. These thermodynamic
energy potentials are determined by the following infintely and non-
linearly coupled integral equations, so called, the thermodynamic
Bethe ansatz equations,

\begin{equation}
\varepsilon_{r,n}(\lambda)+T\sums P_{r,s}^N*\ln\frac{(1+e^{\varepsilon_
{s,n+1}/T})
(1+e^{\varepsilon_{s,n-1}/T})}{(1+e^{\varepsilon_{s-1,n}/T})(1+e^{
\varepsilon_{s+1,n}/T})}
		     =-2\pi JF_r^N(\lambda)\delta_{n,m},
\label{ern}
\end{equation}
where $\varepsilon_{0,n}\equiv\varepsilon_{N,n}\equiv -\infty$ for all
$n$. It is important to notice that Eq. (\ref{den}) and Eq. (\ref{ern})
consist of the same integral kernels and driving terms. This property
makes the zero-field specific heat be obtained analytically at sufficiently
low temperature. We will discuss it in details later.

At T=0 Johannesson \cite{joh} have shown that  only strings of length $m$
contribute. Note that $m$ is nothing but the rank of the tensor
representation of the model. In other words we have only one band ($n=m$)
for each family ($r$) in the ground state. Hence each band is described
by the thermodynamic energy potential $\ep$ which is symmetric and
monotonically
increasing function of $\lambda$. $\ep$ changes sign when symmetry breaking
field $H_r$ is applied. This means that in field $\ep$ crosses zero at a
special value of $\lambda$, designated as $B_r$. Hence the field $H_r$
can be related to $B_r$
defined by $\varepsilon_{r,m}(B_r)=0$. As a consequence of the Fermi
statistics
for the rapidities, $\ep<0$ corresponds to occupied states (particles) and
$\ep>0$ to empty states (holes). Note that $B_r=\infty$ for all $r$ in the
absence of external fields, i.e., $H_r=0$.

In the limit $T \rightarrow 0$ with fields $H_r$ the thermodynamics of the
system is determined by $\ep$ and with some tedious calculation, Eq.
(\ref{ern}) for $n=m$ can be rewritten as the following set of $N-1$
non-linearly coupled integral equations
\begin{equation}
\ep+\sums K_{r,s}^{[N,m]}*T\ln(1+e^{\varepsilon_{s,m}/T})
= \frac{r(N-r)}{N} H_r -2\pi JF_r^N(\lambda), \label{erm}
\end{equation}
where $K_{r,s}^{[N,m]}$ is the Fourier transform of
\begin{equation}
\tilde{K}_{r,s}^{[N,m]}(\omega)= e^{m\frac{|\omega|}{2}}
\frac{\sinh(min(r,s)\frac{\omega}{2})
\sinh([N-max(r,s)]\frac{\omega}{2})}
{\sinh(m \frac{\omega}{2})\sinh(N\frac{\omega}{2})} - \delta_{r,s}.
\label{krs}
\end{equation}
The free energy per site is given by
\begin{equation}
f^{[N,m]}(T,H)=f^{[N,m]}(0,0)-\sums T \intinf F_s^N(\lambda)\ln(1+e^
{\varepsilon_{s,m}/T}), \label{fth}
\end{equation}
where
\[
f^{[N,m]}(0,0)=\frac{2J}{N}\sum_{k=1}^{m} \left[\Psi\left(\frac{k}{N}
\right)-\Psi\left(1+\frac{k-1}{N}\right)\right],
\]
with $\Psi$ being the digamma function.

We consider a level splitting of the $N$-fold multiplets into a ground
multiplet of degeneracy $j$ and an excited $(N-j)$-fold multiplet due to
a small external field $H_j$. This means that $B_j$ is finite, but
all other $B_{r\ne j}=\infty$.
Hence we can rewrite the integral equation (\ref{erm}) for $r=s=j$ as
\begin{equation}
\epj+K_{j,j}^{[N,m]}*T\ln(1+e^{\varepsilon_{j,m}/T})
= \frac{j(N-j)}{N} H_j -2\pi JF_j^N(\lambda).  \label{ejm}
\end{equation}
To evaluate the low-temperature specific heat coefficient in field, it is
sufficient to expand $\epj$ to order $T^2$, i.e., $\varepsilon_{j,m}
\simeq \epsnt + T^2 \epst$.
Using the Sommerfeld formula to expand $\ln(1+e^{\varepsilon_{j,m}/T})$
at low $T$,
we have then the follwing integral equations for $\epsnt$ and $\epst$
\begin{eqnarray}
\epsnt(\lambda)& + &2\int_{B_j}^{\infty} d\lambda K_{j,j}^{[N,m]}
(\lambda-\lambda')
\epsnt(\lambda')=\frac{j(N-j)}{N} H_j -2\pi JF_j^N(\lambda),
\nonumber\\
\epst(\lambda)& + &2\int_{B_j}^{\infty} d\lambda K_{j,j}^{[N,m]}
(\lambda-\lambda') \epst(\lambda') \nonumber \\ [12pt]
& = &\frac{\pi^2}{6}\deps_{B_j}^{-1} [K(\lambda+B_j)+K(\lambda-B_j)].
\label{ejm02}
\end{eqnarray}
The free energy is expressed by
\begin{eqnarray}
f^{[N,m]}(T,H)&=&f(0,0) - 2 \int_{B_j}^{\infty} d\lambda F_j^N(\lambda)
		     \epsnt(\lambda) \nonumber \\[12pt]
        &-&T^2\left(\frac{\pi^2}{3}\deps_{B_j}^{-1} F_j^N(B_j)+2\int_{B_j}^
	       {\infty} d\lambda F_j^N(\lambda) \epst(\lambda)\right).
\label{fth02}
\end{eqnarray}
The integral equations for $\epsnt$  were solved by Schlottmann \cite{sch3}
for very small field yielding the
susceptibility, i.e., Eq. (\ref{sus}) and the following useful relations
\begin{eqnarray}
e^{-\frac{2\pi B_j}{N}} & \simeq & \frac{j(N-j) H_j G_{+}^{(N,m,j)}
(i\frac{2\pi}{N})}{4\pi JG_{+}^{(N,mj)}(0)\sin\frac{j\pi}{N}} Q ,
\label{bjhj}  \\
\deps_{B_j} & = & \frac{2\pi j(N-j) H_j}{N^2G_{+}^{(N,mj)}(0)} Q,
\label{dejm}  \\
Q & = & 1+\frac{1}{2}\left(\frac{m}{2\pi B_j}\right)-
\frac{1}{2}\left(\frac{m}{2\pi B_j}\right)^2\ln\left|\frac{m}{2\pi B_j}
\right|+\cdots \nonumber
\end{eqnarray}
Note that the parameter $B_j$ tends to infinity as $H_j \rightarrow 0$
as expected.

To obtain the in-field specific heat, it is necessary to solve the integral
equations for $\epst$ in (\ref{ejm02}).
Defining $\varphi(\lambda)=\epst(\lambda+B_j)$, the integral equation
becomes
\begin{eqnarray}
\varphi(\lambda)&+&\int_0^{\infty}d\lambda'[K(\lambda-\lambda')+K(\lambda
+\lambda'+B_j)]\varphi(\lambda')\nonumber \\
&=&\frac{\pi^2}{6}\deps_{B_j}^{-1}[K(\lambda)+K(\lambda+2B_j)].\nonumber
\end{eqnarray}
This equation can be solved iteratively for large $B_j$ since the integral
kernel $K(B_j)\sim\frac{1}{B_j}$.
Hence substituting $\varphi \simeq \varphi_1 + \varphi_2 + \cdots$,
\begin{eqnarray}
\varphi_1(\lambda)+\int_0^{\infty}d\lambda'K(\lambda-\lambda')\varphi_1
(\lambda')
&=&\frac{\pi^2}{6}\deps_{B_j}^{-1}K(\lambda), \label{phi1}  \\
\varphi_2(\lambda)+\int_0^{\infty}d\lambda'K(\lambda-\lambda')\varphi_2
(\lambda')
&=&\varphi_1(-\lambda-2B_j).\nonumber
\end{eqnarray}
These equations are of the Wiener-Hopf type and can be solved by standard
methods, yielding for $\lambda\geq 0$,
\begin{equation}
\varphi(\lambda)=\intinf\frac{d\omega}{2\pi}e^{-i\omega\lambda}
		 \left[\frac{\pi^2}{6}\deps_{B_j}^{-1}
\left(\frac{1}{G_{+}^{(N,m,j)}(\omega)}-1\right)\right], \label{phi} \\
\end{equation}
where $G_{+}^{(N,m,j)}(\omega)$ is defined in Eq. (\ref{g+}). This solution
does not have $\varphi_2$ contributions since $\varphi_2$ is of order
$\frac{1}{B_j^2}$ or higher and we are interested only in the leading
contribution to the in-field $\gam$. Collecting the results and inserting
those into the free energy (\ref{fth02}), we straightforwardly obtain that
\[
\lim_{H_j\rightarrow 0}\limt\gam_j = \frac{N}{12J}[1+G_{+}^{(N,m,j)}
(i\frac{2\pi}{N})]
\]
For an arbitrary level splitting the linear superposition principle can be
used as long as all $H_j$ are small. In particular for a Zeeman
splitting $H_j=H$,
\begin{equation}
\limh\limt\gam = \gamh  \label{gamh}
\end{equation}

Now let us consider the case when $0<T\ll J$ and $H_j=0$. In this
case we can calculate the entropy by the method used by Babujian \cite{bab}
and Filyov et. al. \cite{ftw}. As mentioned before, all rapidities obey
Fermi statistics. This implies that the entropy per site of the model can
be written as
\begin{equation}
\it{S}=\sums\sum_{n=1}^{\infty}\intinf d\lambda[(\sigma_{s,n}
+\sigma_{s,n}^h)
\ln(\sigma_{s,n}+\sigma_{s,n}^h)
-\sigma_{s,n}\ln\sigma_{s,n}-\sigma_{s,n}^h\ln\sigma_{s,n}^h],
\label{ent}
\end{equation}
where the densities of occupied ($\sigma_{s,n}$) and unoccupied (
$\sigma_{s,n}^h$) states satisfy the integral equations (\ref{den}).
To calculate the entropy, we analyze the two sets of integral equations
(\ref{den}) and (\ref{ern}) which have the same integral kernels $P_{r,s}^N$
and driving terms $F_r^N$ as mentioned before. Substituting
$\lambda\rightarrow\Lambda-\frac{N}{2\pi}\ln\frac{NT}{2\pi}$ in Eqs.
(\ref{den})
and (\ref{ern}) and differenciating Eq.(\ref{ern}) with respect to
$\Lambda$,
we have
the interesting relations between the energy potentials and the densities
of states, i.e.,
\begin{equation}
\sigma_{r,n}(\Lambda)=\frac{N}{4\pi^2J}\frac{d\varepsilon_{r,n}}{d\Lambda}
f(\varepsilon_{r,n}), \,\,
\sigma_{r,n}^h(\Lambda)=\frac{N}{4\pi^2J}\frac{d\varepsilon_{r,n}}
{d\Lambda}[1-f(\varepsilon_{r,n})],
\label{denern}
\end{equation}
where $f(x)=(1+e^{x/T})^{-1}$. Note that these relations are valid only for
sufficiently low $T$ and
$[0,\infty]$ in $\lambda$-space is mapped on $[-\infty,\infty]$ in
$\Lambda$-space. For the convinience let us define $E_{r,n}(\Lambda)
=\varepsilon_{r,n}(\lambda-\frac{N}{2\pi}\ln\frac{NT}{2\pi})/T$ and use
the relation (\ref{denern}) to rewrite the entropy (\ref{ent}) as
\begin{eqnarray}
\it{S}=-\frac{NT}{2\pi^2J}\sums\sum_{n=1}^{\infty}
\int_{E_{s,n}^{min}}^{E_{s,n}^{max}}
&&dE_{s,n}[f(TE_{s,n})\ln f(TE_{s,n})\nonumber \\
&&+(1-f(TE_{s,n}))\ln(1-f(TE_{s,n}))],
\label{ent1}
\end{eqnarray}
where $E_{r,n}$ satisfies the following integral equations
\[
E_{r,n}(\Lambda)-\sums P_{r,s}^N*\ln\frac{(1+e^{E_{s,n+1}})
(1+e^{E_{s,n-1}})}
{(1+e^{E_{s-1,n}})(1+e^{E_{s+1,n}})}=-J\sin(\frac{r}{N}\pi)e^
{-\frac{2\pi}{N}\Lambda}\delta_{n,m}
\]
with the asymptotic condition $\lim_{n\rightarrow\infty}\frac
{E_{r,n}}{n}=0$. These equations imply that $E_{r,n}^{max}=E_{r,n}
(\infty)$ and $E_{r,n}^{min}=E_{r,n}(-\infty)$.
Hence the integral equations for maximum or mimimum solutions are reduced
to the difference equations such that
\begin{eqnarray}
E_{r,n}^{max}&=&\sums \frac{min(r,s)}{N}[N-max(r,s)]\ln\frac{(1+e^
{E_{s,n+1}^{max}})(1+e^{E_{s,n-1}^{max}})}
{(1+e^{E_{s-1,n}^{max}})(1+e^{E_{s+1,n}^{max}})}, \nonumber \\
E_{r,n\ne m}^{min}&=&\sums \frac{min(r,s)}{N}[N-max(r,s)]\ln\frac{(1+e^
{E_{s,n+1}^{min}})(1+e^{E_{s,n-1}^{max}})}
{(1+e^{E_{s-1,n}^{min}})(1+e^{E_{s+1,n}^{min}})}  \nonumber \\
E_m^{min}&=&-\infty .\nonumber
\end{eqnarray}
The solutions of these equations are of the form,
\begin{eqnarray}
E_{r,n}^{max}&=&\ln\left[\frac{(r+n)(N-r+n)}{r(N-r)}-1\right], \nonumber \\
E_{r,n<m}^{min}&=&\ln\left[\frac{\sin(\frac{r+n}{m+N}\pi)\sin(
\frac{N-r+n}{m+N}\pi)}{\sin(\frac{r}{m+N}\pi)\sin(\frac{N-r}{m+N}\pi)}
-1 \right],
\label{sde} \\
E_{r,n \ge m}^{min}&=&\ln\left[\frac{(r+n-m)(N-r+n-m)}{r(N-r)}-1\right].
\nonumber
\end{eqnarray}
Note that the solutions for $N=2$ with arbitrary $m$ were obtained by
Babujian \cite{bab} who has treated the $SU(2)$ invariant Heisenberg chain
with arbitrary spin $S$. The above solutions (\ref{sde}) correctly reproduce
those for $N=2$ with $m=2S$.

Now we can express the entropy (\ref{ent1}) as following, noticing that
$E_{r,n}^{max}=E_{r,n+m}^{min}$,
\begin{equation}
\it{S}=-\frac{NT}{2\pi^2J}\sum_{r=1}^{N-1}\sum_{n=1}^{m}
\int_{0}^{x_{r,n}}dx\left[\frac{\ln x}{1-x}+\frac
{\ln(1-x)}{x}\right],
\label{ent2}
\end{equation}
where
\[
x_{r,n}=\frac{\sin(\frac{r}{m+N}\pi)\sin(\frac{N-r}{m+N}\pi)}
{\sin(\frac{r+n}{m+N}\pi)\sin(\frac{N-r+n}{m+N}\pi)}.
\]
We evaluate numerically the sums of the dilogarithmic functions,
yielding
\begin{equation}
-\sum_{r=1}^{N-1}\sum_{n=1}^{m}
\int_{0}^{x_{r,n}}dx\left[\frac{\ln x}{1-x}+\frac{\ln(1-x)}{x}\right]
=\frac{\pi^2}{3}\frac{m(N^2-1)}{m+N}.
\label{sum}
\end{equation}
Note that this sum rule is exact even though we have obtained it
numerically. Hence the zero-field specific heat coefficient
\[
\limt\limh\gam=\gamt,
\]
which agrees with the result from the conformal field theory. It states
that the linear specific heat coefficient $\gamma=\frac{c\pi}{3v}$
\cite{aff}, where the central charge $c=m(N^2-1)/(m+N)$ \cite{vega,am}
and the group velocity $v=\frac{2\pi J}{N}$ \cite{sut2} for SU(N)
Heisenberg antiferromagnet. Note that for a sufficiently low temperature
the all excitations have the same group velocities.

The $\gam$ in finite field is less than the one in zero field as expected
since the entropy is reduced in the ordered system. This result also
reproduces the previous results for SU(2) invariant antiferromagnetic
Heisenberg chain \cite{ls3,lee}, noticing that the coupling
constant $J$ differs by a factor 2.

We would like to thank Dr. C. Ahn for fruitful discussions and also to
acknowledge support by NON DIRECTED RESEARCH FUND, Korea Research
Foundation, 1993.

% This is a file containing the references.
%\newpage

\end{document}